# Hierarchical Approach for Online Mining –Emphasis towards Software Metrics

M .V.VIJAYA SARADHI

Dept. of Comp. Sc & Eng.

ASTRA

Hyderabad, India.

B.R.SASTRY

Director

ASTRA

Hyderabad, India

P.SATISH

Dept. of CSE

VIE

Hyderabad, India

**Abstract ----Several multi-pass algorithms have been proposed for Association Rule Mining from static repositories. However, such algorithms are incapable of online processing of transaction streams. In this paper we introduce an efficient single-pass algorithm for mining association rules, given a hierarchical classification amongest items. Processing efficiency is achieved by utilizing two optimizations, hierarchy aware counting and transaction reduction, which become possible in the context of hierarchical classification.**

**This paper considers the problem of integrating constraints that are Boolean expression over the presence or absence of items into the association discovery algorithm. This paper present three integrated algorithms for mining association rules with item constraints and discuss their tradeoffs. It is concluded that the variation of complexity depends on the measure of DIT (Depth of Inheritance Tree) and NOC (Number of Children) in the context of Hierarchical Classification.**

## I. INTRODUCTION

The aim of Association Rule Mining is to find latent associations among data entities in database repositories, a typical example of which is the transaction database maintained by a supermarket. An association rule is an implication of the form A => B, which conveys that customers buying set of items A would also with a high probability buy set of items B. The concept of association rule mining was first introduced in [4]. Typically the problem is decomposed into two phases. Phase I of the problem involves in finding the frequent item sets in the database, based on a pre- defined frequency threshold minsupport. Phase II of the problem involves generating the association rules from the frequent item sets found in Phase I. Typically, the reported approaches in Phase I re- quire multiple passes over the transaction database to determine the frequent item sets of deferent lengths [1, 2, 3]. All these approaches assume that a static database is available, so that multiple scans can be made over it. With online systems, it is desirable to make decisions on the fly, processing data-streams in- stead of stored databases.

In this paper, we aim at a online algorithm, capable of processing online streams of transactions. Assume that the algorithm has computed its result up to and including the first *n* transactions. A true online algorithm should be capable of updating the result for the (*n* + 1). The transaction, without requiring a re-scan over the past *n* transactions. In this way such an algorithm can handle transaction streams. In fact it is true that items in an online shopping mart or a supermarket are categorized into sub-classes, which in turn make up classes at a higher level, and so on. Besides the usual rules that involve individual items, learning association rules at a particular sub- class or class level is also of much potential use and significance, e.g. an item-specific rule such as Customers buying Brand A sports shoes tend to buy Brand B tee- shirts" may be of less practical use than a more general rule such as Customers buying sports shoes tend to buy tee-shirts". With this aim, we can be made of commonly employed hierarchical classification of items to devise a simple and efficient rule mining algorithm. [2] Proposes a single-pass algorithm for hierarchical online association rule mining.

In this paper, we refer to this algorithm as HORM. The present work carries forward the idea of [1], and proposes an efficient algorithm for Phase I. The present work also looks at Phase II, i.e. the generation of association rules. [9] Proposes an algorithm to generate non-redundant rules. We present a modified algorithm for Phase II that better suits the need to mine hierarchical association rules. For example, they may only want rules that contain a specific item or rules that contain children of a specific item in a hierarchy. While such constraints can be applied as a post processing step, integrating them into the mining algorithm can dramatically reduce the execution time. In practice, users are often interested only in a sub set of associations, for instance, those containing at least one item from a user-defined subset of items. When taxonomies are present, this set of items may be specified using the taxonomy, e.g. all descendants of a given item. While the output of current algorithms can be filtered out in a post-processing step, it is much more efficient to incorporate such constraints into the association's discovery algorithm.

Design choices on the hierarchy employed to represent the application are essentially choices about restricting or expanding the scope of properties of the classes of objects in the application. Two design decisions which relate to the inheritance hierarchy can be defined [11]. They are depth of inheritance (DIT) of a class and the number of children of the class (NOC).Depth of Inheritance of the class is the DIT metric for the class. The DIT (Depth of Inheritance Tree) will be the maximum length from the node to the root of the tree. The deeper a class is in the hierarchy, the greater the number of methods it is likely to inherit, making it more complex to predict its behavior. The Deeper trees constitute greater design complexity. The deeper a particular class is in the hierarchy, the greater the potential reuse of inherited methods. The inheritance hierarchy is a directed acyclic graph, which can be described as a tree structure with classes as nodes, leaves and a root.

The NOC (Number of Children) is the number of immediate subclasses subordinated to a class in the class hierarchy. It is a measure of how many subclasses are going to inherit the methods of the parent class. The number of children gives an idea of the potential influence a class has on the design. If a class has a large number of children, it may require more testing of the methods in that class. The Greater the number of children (NOC), greater the likelihood of improper abstraction of the parent class. If a class has a large number of children; it may be a case of misuse of sub classing. In this paper, we consider constraints that are boolean expressions over the presence or absence of items in the rules. When taxonomies are present, we allow the elements of the boolean expression to be of the form ancestors (item) or descendants (item) rather than just a single item.

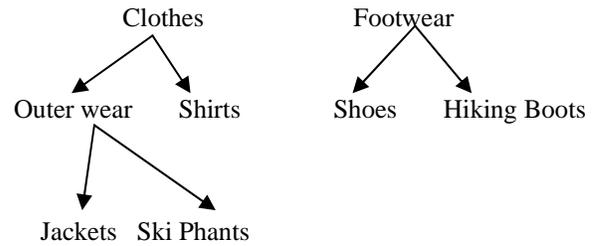

Figure 1.  Example for taxonomy

For example,(Jacket A Shoes) V (descendants (Clothes) A7 ancestors (Hiking Boots)) expresses the constraint that we want any rules that either (a) contain both Jackets and Shoes, or (b) contain Clothes or any descendants of clothes and do not contain Hiking Boots or Footwear.

II.     THEORY

A. Basic Concepts and Problem Formulation

Hierarchical classification of data means that the items which make up a transaction are categorized into Classes, sub-classes, and so on. While doing Hierarchical Classification of data, some measures to be consider. Design of a class involves decisions on the scope of the methods declared within the class. We have to consider four major features in the Hierarchical classification of data in terms of Classification tree 1.Identification of classes.2.Identify the semantics of classes.3.Identify relations between classes.4.Implementation of classes. Using several metrics can help designers, who may be unable to review design complexity for the entire application [11]. The Depth of inheritance Tree (DIT) and Number of children (NOC) metrics check whether the application is getting too heavy (i.e .too many classes at the root level declaring many methods).Classes with high values of DIT tend to complex classes. Evidently it is possible to mine for two types of rules: an item-specific rule such as Customers buying soap of brand A tend to buy canned soup of brand B", or a more general Rule such as Customers buying soaps tend buy canned soup". The latter is an association on classes or sub-classes, rather than on individual items. Let $I$ be the set of all items stored in, say, a typical supermarket.

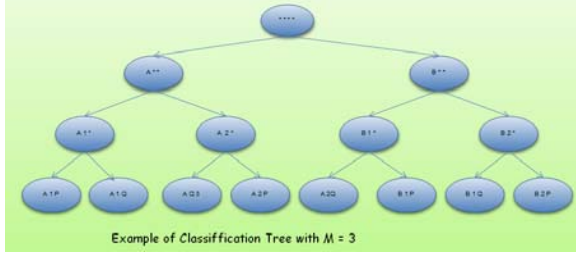

Figure 2. Example of Classification Tree

We suppose that, at each level of classification, a fixed number $M$ of classes, sub-classes or items are present. At the root level we have classes $C1; C2; C3: CM$. At the next level, for a class $Ck$, we will have the $M$ sub-classes $Ck1;Ck2 : : :CkM$. For $jIj = 20000$, and with $M = 12$, for example, we will need four levels of classification; the last level will contain individual items stored in transaction, which will be coded as $Cjklm$, i.e. one index for each level of classification. A hierarchical association rule is an implication of the type $X => Y$ where $X; Y$ are disjoint subsets of the sub-classes of some $C\alpha$, the parent class of $X$ and $Y$. As usual, support for association rule $X =>Y$ is defined as the fraction of transactions in the transaction database which contain $X=>Y$; confidence of the rule is defined as the fraction of transactions containing $X$ which also contain $Y$. We denote the support and confidence of rule $X => Y$ as $supp(X=> Y)$ and $conf(X => Y)$ respectively.

We may also write $XY$ to represent $X => Y$. Subsets of sub-classes of class $C$ are elements of the power-set of the set of sub-classes of $C$. For a given class $C$, the counts of all subsets occurring in the transaction database are stored in an integer array, called count array, of size $2M$. The natural bitmap representation of a subset can be used directly as the index of the corresponding cell in the count array.

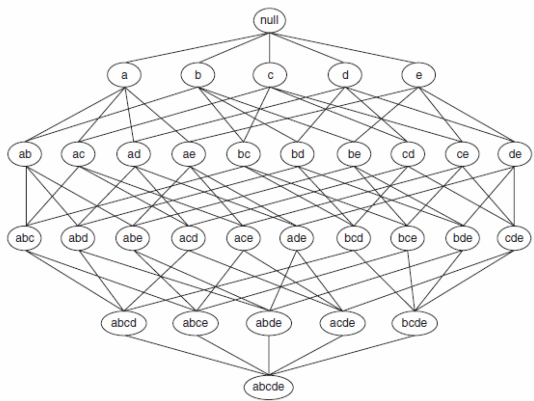

Figure 3: Item set Lattice

**Phase 1:**

Find all frequent item sets (item sets whose support is greater than minimum support) that satisfy the boolean expression B. Recall that there are two types of operations used for this problem: candidate generation and counting support. The techniques for counting the support of candidates remain unchanged. However, as mentioned above, the A priori candidate generation procedure will no longer generate all the potentially frequent itemsets as candidates when item constraints are present. We consider three different approaches to this problem.

- Generate a set of selected items S such that any Item sets that satisfy B will contain at least one selected item.

- Modify the candidate generation procedure to only count candidates that contain selected items.

- Discard frequent item sets that do not satisfy B. The third approach, "Direct" directly uses the boolean expression B to modify the candidate generation procedure so that only candidates that satisfy B are counted (Section 4.2).

**Phase 2:**

To generate rules from these frequent itemsets, we also need to find the support of all subsets of frequent itemsets that do not satisfy?. Recall that to generate a rule AI3, we need the support of AB to find the confidence of the rule. However, AB may not satisfy B and hence may not have been counted in Phase 1. So we generate all subsets of the frequent itemsets found in Phase 1, and then make an extra pass over the dataset to count the support of those subsets that are not present in the output of Phase 1.

**Phase 3:**

Generate rules from the frequent item sets found in Phase 1, using the frequent item sets found in Phases 1 and 2 to compute confidences, as in the A priori algorithm. We discuss next the techniques for finding frequent item sets that satisfy Z? (Phase 1). The algorithms use the notation in Figure 2.

**B. Approaches using Selected Items**

Generating Selected Items Recall the boolean expression Z? = D1 V D2 V . . . V D,,,,, where Di = ail A Q~Z A u . e A ain; and each element oij is either lij or dij, for some Zij E C. We want to generate a set of items S such that any item set that satisfies Z? will contain at least one item from S. For example, let the set of items {l,2,3,4,5}.

Consider B=(lA2)V3.The sets (1, 3}, {2, 3) and (1, 2, 3, 4, 5) all have the property that any (non-empty) item set that satisfies B will contain an item from this set. If B = (1 A 2) V 73, the set (1, 2, 4, 5) has this property. Note that the inverse does not hold: there are many itemsets that contain an item from S but do not satisfy B. For a given expression B, there may be many different sets S such that any itemsets that satisfies B contains an item from S. We would like to choose a set of items for S so that the sum of the supports of items in S is minimized. The intuition is that the sum of the supports of the items is correlated with the sum of the supports of the frequent itemsets that contain these items, which is correlated with the execution time. We now show that we can generate S by choosing one element oij from each disjunct 0; in B, and adding either lij or all the elements in .C - {Zij} to S, based on .whether oij is lij or li.j respectively

### C. Hierarchical representation:

The traditional model of a relational database is a set of relations, which cannot directly encode hierarchical properties among data, such as the is-a relationship. To address this issue, we represented both theses relational and hierarchical properties of data explicitly within a mining ontology. Each domain class in the ontology, called a Node, corresponds to a relation in the database. The subclass hierarchy under Node encodes an "is-a" relationship among domain classes. Each class has properties that contain string values mapping to the column names of the database table that stores the instances for that class. The use of Ontology to encode an explicit representation of data allows reuse of the mining method with different database schemas and domains, since such encoded knowledge can be easily modified. The mining ontology serves as a bridge between the database and the mining algorithm, and guides the hierarchical search of the latter across multiple tables within the former.

### III. Mining Algorithm

This data mining approach undertakes rule association analysis between two input domain classes and their subclasses in the mining ontology. Standard rule association mining looks for frequently occurring associations between input values that meet the minimal criteria of user defined *interestingness*, such as confidence (the probability of one value occurring given another) and Support (the probability of two values occurring together). The Chrono Miner algorithm extends this standard approach by also examining the occurrence of different temporal relationships between the time stamps of those values.

```
MineTemporalAssociation (class1, class2,
temporalPatternVector[], min_support, min_confidence) {

Vector temporalAssociationRules
for every childNodes of class1 { /*traverse the subtree of
class1 (breadth first search)*/
  for every childNodes of class2 { /*traverse the subtree of
class2 (breadth first search) */
  inputSet = Join data of child (class1) and child (class2)
  if (count(inputSet) > min_support) {
    for (every inputSet) {
      for every temporalPattern in temporalPatternVector {
        if (temporalPattern (child (class1), child (class2)) {
          Rule candidateRule = Rule (class1.value, class2.value,
temporalPattern)
          if (temporalAssociationRules.find (candidateRule ))
            Increment the count
          else
            temporalAssociationRules.Add (candidateRule)
        } /* if ends for candidate association rule */
      } /* for loop ends for every temporal patterns */
    } /* for loop ends for every inputSet */
  } /* if ends for input set greater than min_support */
} /* for loop ends for childNodes of class2 */
} /* for loop ends for childNodes of class1 */

For every temporalAssociationRules {
  If ((temporalAssociationRules.min_confidence () <
confidence)
    || (temporalAssociationRules.support () < min_support))
    Delete that rule
  }
return temporalAssociationRules
```

**Temporal Association Rule algorithm**

Using the mining ontology, the search for temporal associations involves partial or complete traversal of the hierarchical structure starting from each input class, proceeding through top-down induction as described in the pseudo code presented in Figure 4.

### A. HORM Algorithm

From the classes and sub-classes which make up the classification tree, the user selects a *set of classes of interest* [4], to be denoted here as *SIC*. Association rules to be mined are of the type $X \Rightarrow Y$ where $X$ and $Y$ are disjoint subsets of a class or sub-class of interest. The problem of hierarchical association rule mining is now defined as: Find all association rules of the type $X \Rightarrow Y$, within each class of interest in *SIC*, which have a specified minimum support and confidence in the transaction database or stream. To find associations within a class or sub-class of interest, we need to maintain counts for all its subsets. For the class $A^{***}$, with $M = 4$, for example, we need to count the occurrences in the transaction database of all the subsets of $\{A1^{***}A2^{***}\ A3^{***}A4^{***}\}$. Clearly there are $2M - 1$ non-empty subset combinations for a sub-class with $M$ elements. Therefore a count array of size $2M - 1$ needs to be maintained for each class or sub-class of interest. HORM algorithm takes as input the i.e. support values, in the transaction database of all the subsets of the classes or sub-classes of interest. The time complexity of this algorithm is $O(|D|K2M)$ [6]. The memory requirement of HORM is $K2M$, since each element of *SIC* requires an array of size $2M$.

## IV. Enhancements Proposed

### A. Hierarchy-Aware Counting

In HORM, each transaction is checked against all the classes or sub-classes in *SIC*. But suppose we have two classes or sub-classes in *SIC* of which one is itself a sub-class of the other. In HORM, the per-transaction code is executed once for each of these elements of *SIC*, without taking into account this hierarchical relationship between the two. But clearly if the first iteration suggests that the current transaction does not support, say $PQ^{**}$, we do not need to iterate for any of its sub-class such as $PQR^*$. We apply this intuition to speed up the algorithm: If a transaction does not support a class or sub-class,
it does not support any of its sub-classes either. We call this first enhancement *hierarchy-aware counting*.

### B. Transaction Reduction

This second enhancement reduces the computation within the inner loop. For every class or sub-class in *SIC*, HORM processes the current transaction in its entirety. However, suppose we have two classes or sub-classes in *SIC* which do not share an ancestor-descendant relationship. Once we have matched the entire transaction against the first class or sub-class clearly it is not necessary to again match the entire transaction against the second one as well. Suppose $A^{***}$ and $B^{***}$ are two classes of interest, and let the current transaction $T$ be $\{A1Q6;A2P6;B2Q6;B1Q7;A2P7;B2P7\}$. While $T$ is being checked against $A^{***}$, the algorithm in fact traverses through the items of $T$ and finds the sub-transaction $T=A^{***} = \{A1Q6;A2P6;A2P7\}$, which may be called the *projection* of class $A^{***}$ on $T$. Clearly $T=A^{***}$ does not contain any items that belong to $B^{***}$, because the sub-classes of $A^{***}$ and $B^{***}$ are disjoint. Thus we can remove $T=A^{***}$ from $T$ and pass the remaining items $T1 = T ¡ T=A^{***}$ to match against $B^{***}$. Thus the part of a transaction that is a projection of a class can be removed to obtain a reduced transaction to match against disjoint classes. We call this second enhancement *transaction reduction*.

### C. Non-Redundant Rule Generation

The version implemented in the present work is based on the basic concepts proposed in [5]. The hierarchical rule mining technique described here does not require a separate adjacency lattice of the classes or subclasses of interest. The count arrays described above can themselves be viewed as adjacency lattices used in [5], leading to very clean design and implementation.

## CONCLUSIONS

This proposed algorithm, Modified Hierarchical online rule mining, or MHORM, which optimizes the time requirements of the earlier reported algorithm HORM [8].

We considered the problem of discovering association rules in the presence of constraints that are Boolean expressions over the presence of absence of items. Such constraints allow users to specify that they are interested in we presented three such integrated algorithm, and discussed the tradeoffs between them. Empirical evaluation of the Multiple Joins algorithm on three real-life datasets showed that integrating item constraints can speed up the algorithm by a factor of 5 to 20 for item constraints with selectivity between 0.1 and 0.01.For candidates that were not frequent in the sample but were frequent in the datasets, only those extensions of such candidates that satisfied those constraints would be counted in the additional pass.

It is concluded that while constructing Classification Tree, the measure of Depth of Inheritance Tree (DIT) with respect to Number of children (NOC) place a dominant role, which is evidence from the fact that the complexity depends on the depth of inheritance Tree (DIT) with respect to Number of children (NOC). Both DIT and NOC directly relate to the layout of the class hierarchy. In an Classification Tree, Classes with high DIT values are associated with a higher number of defects.

## AUTHORS PROFILE

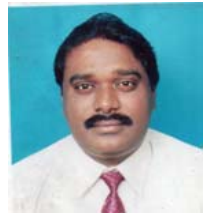

**M.V.Vijaya Saradhi** is Currently Associated Professor in the Department of Computer Science and Engineering (CSE) at Aurora's Scientific, Technological and Research Academy, (ASTRA), Bandlaguda, Hyderabad, India, where he teaches Several Courses in the area of Computer Science. He is Currently Pursuing the PhD degree in Computer Science at Osmania University, Faculty of Engineering, Hyderabad, India. His main research interests are Software Metrics, Distributed Systems, Object-Oriented Modeling (UML), Object-Oriented Software Engineering, Data Mining, Design Patterns, Object- Oriented Design Measurements and Empirical Software Engineering. He is a life member of various professional bodies like MIETE, MCSI, MIE, MISTE. **E-mail: meduri_vsd@yahoo.co.in**

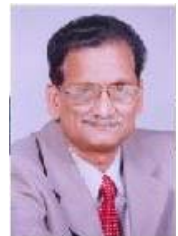

**Dr. B. R. Sastry** is currently working as Director, Astra, Hyderabad, India. He earlier worked for 12 years in Industry that developed indigenous computer systems in India. His areas of research includes Computer Architecture, Network Security, Software Engineering, Data Mining and Natural Language Processing, He is currently concentrating on improving academic standards and imparting quality engineering

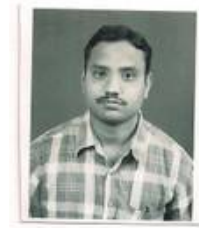

P. Satish is Currently Asst Professor in Department of computer Science & Engineering at Vivekananda Institute of Engineering (VIE), Hyderabad, India.